\documentclass[cup7b]{cupbook}

\begin{document}
%


\pagenumbering{roman}




\def\Box{{\hbox{$\sqcup$}\llap{\hbox{$\sqcap$}}}}

\def \lsim{\mathrel{\vcenter
     {\hbox{$<$}\nointerlineskip\hbox{$\sim$}}}}
\def \gsim{\mathrel{\vcenter
     {\hbox{$>$}\nointerlineskip\hbox{$\sim$}}}}

\cleardoublepage
\pagenumbering{arabic}

\author[C.P. Burgess]{C.P. BURGESS\\Department of Physics \& Astronomy,
        McMaster University,\\ 1280 Main St. W,
        Hamilton, Ontario, Canada, L8S 4M1\\ {\it and} \\
        Perimeter Institute for Theoretical Physics,\\
        31 Caroline St. N,
        Waterloo, Ontario, Canada, N2L 2Y5.}

\chapter{Effective Theories and Modifications of Gravity} 

\begin{abstract}
We live at a time of contradictory messages about how successfully
we understand gravity. General Relativity seems to work very well
in the Earth's immediate neighborhood, but arguments abound that
it needs modification at very small and/or very large distances.
This essay tries to put this discussion into the broader context
of similar situations in other areas of physics, and summarizes
some of the lessons which our good understanding of gravity in the
solar system has for proponents for its modification over very
long and very short distances. The main message is that effective
theories, in the technical sense of `effective', provide the
natural language for testing proposals, and so are also effective
in the colloquial sense.
\end{abstract}

\section{Introduction}

Einstein's recognition early last century that gravity can be
interpreted as the curvature of space and time represented an
enormous step forward in the way we think about fundamental
physics. Besides its obvious impact for understanding gravity over
astrophysical distances --- complete with resolutions of earlier
puzzles (like the detailed properties of Mercury's orbit) and
novel predictions for new phenomena (like the bending of light and
the slowing of clocks by gravitational fields) --- its
implications for other branches of physics have been equally
profound.

These implications include many ideas we nowadays take for
granted. One such is the universal association of fundamental
degrees of freedom with fields (first identified for
electromagnetism, but then cemented with its extension to gravity,
together with the universal relativistic rejection of action at a
distance). Another is the recognition of the power of symmetries
in the framing of physical law, and the ubiquity in particular of
gauge symmetries in their description (again reinforcing the
earlier discovery in electromagnetism). A third is the
systematization of the belief that the physical content Nature's
laws should be independent of the variables used in their
description, and the consequent widespread penetration of
geometrical methods throughout physics.

But the study of General Relativity (GR) and other interactions
(like electromagnetism, and its later-discovered relatives: the
weak and strong forces) have since drifted apart. Like ex-lovers
who remain friends, for most of the last century practitioners in
either area have known little of the nitty gritty of each other's
day-to-day struggles, even as they read approvingly of their
occasional triumphs in the popular press.

Over the years the study of both gravity and the other
interactions has matured into precision science, with many
impressive theoretical developments and observational tests. For
gravity this includes remarkably accurate accounts of motion
within the solar system, to the point that GR --- through its use
within the global positioning system (GPS) --- is now an
indispensable tool for engineers [Will 2001]. For the other
interactions the successes include the development and testing of
the Standard Model (SM), a unified framework for all known
non-gravitational physics, building on the earlier successes of
Quantum Electrodynamics (QED).

There is nevertheless a mounting chorus of calls for modifying
General Relativity, both at very short and very long distances.
These arise due to perceived failures of the theory when applied
over distances much different from those over which it is
well-tested. The failures at short distances are conceptual, to do
with combining gravity with quantum effects. Those at long
distances are instead observational, and usually arise as ways to
avoid the necessity for introducing the dark matter or dark energy
that seem to be required when General Relativity is applied to
describe the properties of the universe as a whole.

The remainder of this chapter argues that when searching for
replacements for GR over short and long distances there is much to
be learned from other branches of physics, where similar searches
have revealed general constraints on how physics at different
scales can relate to one another. The hard-won lessons learned
there also have implications for gravitational physics, and this
recognition is beginning to re-establish the connections between
the gravitational and non-gravitational research communities.

In a nutshell, the lessons distilled from other areas of physics
make it likely that it is much more difficult to modify gravity
over very long distances than over very tiny ones. This is because
very broad principles (like unitarity and stability) strongly
restrict what is possible. The difficulty of modifying gravity
over long distances is a very useful (but often neglected) clue
when interpreting cosmological data, because it strongly
constrains the theoretical options that are available. We ignore
such clues at our peril.

This chapter is also meant to be colloquial rather than
authoritative, and so citations are not thorough. My apologies to
those whose work is not properly cited.

\section{Modifying Gravity over Short Distances}

The demand to replace General Relativity at short distances arises
because quantum mechanics should make it impossible to have a
spacetime description of geometry for arbitrarily small scales.
For example, an accurate measurement of a geometry's curvature,
$R$, requires positions to be measured with an accuracy, $\delta$,
smaller than the radius of curvature:
\begin{equation} \label{deltacriterion}
 \delta^2 < 1/R \,.
\end{equation}
But for position measurements with resolution, $\delta$, the
uncertainty principle requires a momentum uncertainty, $p \simeq
\hbar/\delta$, which implies an associated energy uncertainty, $E
\simeq p\, c \simeq \hbar c/\delta$, or equivalently a mass $M
\simeq E/c^2 \simeq \hbar/\delta c$. But the curvature associated
with having this much energy within a distance of order $\delta$
is then $R \simeq GM/\delta^3 c^2 \simeq G\hbar/\delta^4 c^3 =
\ell_p^2/\delta^4$, where $\ell_p$ defines the Planck length,
$\ell_p^2 = G \hbar/c^3$, and $G$ is Newton's constant. Requiring
eq.~(\ref{deltacriterion}), then shows that there is a lower bound
on the resolution with which spacetime can be measured:
\begin{equation}
 \delta > \ell_p \simeq \sqrt{ \frac{G \hbar}{c^3} }
 \simeq 1.6 \times 10^{-35} \; \hbox{m} \,.
\end{equation}
Although this is an extremely short distance (present experiments
only reach down to about $10^{-19}$ m), it is also only a lower
bound. Depending on how gravity really works over short distances,
quantum gravity effects could arise at much longer scales.

Notice how crucial it is to this argument that the interaction
strength, $G$, has dimensions of length (in fundamental units, for
which $\hbar = c = 1$). Imagine performing a similar estimate for
an electrostatic field. The Coulomb interaction energy between two
electrons separated by a distance $\delta$ is $E_c \simeq
e^2/\delta$, where $q = -e$ denotes the electron's electric
charge. But the energy required by the uncertainty principle to
localize electrons this close to one another is $E \simeq \hbar
c/\delta$, so the condition that this be smaller than $E_c$ is
\begin{equation}
 \alpha = \frac{e^2}{4\pi \hbar c} < 1 \,,
\end{equation}
where the fine-structure constant, $\alpha \simeq 1/137$, is
dimensionless. This condition doesn't depend on $\delta$ because
the relative strength of quantum fluctuations to electrostatic
interactions does not change with distance.

\subsection{Gravity and renormalizability}

The observation that quantum fluctuations do not get worse at
shorter distances in electrodynamics\footnote{There is a sense in
which quantum effects in QED do get worse at smaller distances,
because the theory is not asymptotically free. But this problem
only arises logarithmically in $\delta$, and so is much less
severe than the power-law competition found above for gravity.}
but do for gravity can be more technically expressed as the
statement that QED is a {\em renormalizable} quantum field theory
(QFT) while GR is not. In QFT small-distance quantum fluctuations
appear (within perturbation theory) as divergences at small
distances (or high momenta) when summing over all possible quantum
intermediate states.

For instance, given a Hamiltonian, $H = H_0 + H_{\rm int}$, the
second-order shift in the energy of a state $|n\rangle$ is
\begin{equation}
 \delta_2 E_n = \sum_m \frac{| \langle n | H_{\rm int} | m \rangle
 |^2}{E_m - E_n} \simeq \int \frac{{\rm d}^3 {\bf p}}{(2\pi)^3}
 \; \frac{| \langle n | H_{\rm int} | {\bf p} \rangle
 |^2}{E({\bf p}) - E_n} + \cdots\,,
\end{equation}
where the approximate equality focusses on the sum over a basis of
free single-particle states having energies $E({\bf p}) =
\sqrt{{\bf p}^2 + m^2}$ when performing the sum over $|m \rangle$.
Because the combination $|\langle n | H_{\rm int} | {\bf p}
\rangle|^2/[E({\bf p}) - E_n]$ typically falls with large $p =
|{\bf p}|$ like $1/p^3$ or slower, the integration over the
momentum of the intermediate state diverges in the ultraviolet
(UV), $p \to \infty$, limit. (Relativistic calculations organize
these sums differently to preserve manifest Lorentz invariance at
each step, but the upshot is the same.)

Renormalizability means that these divergences can all be absorbed
into the unknown parameters of the theory --- like the electron's
charge and mass, for instance --- whose values must in any case be
inferred by comparison with experiments. As the above estimates
suggest, the hallmark of a nonrenormalizable theory is the
appearance of couplings (like Newton's constant) having dimensions
of length to a positive power (in fundamental units). Couplings
like this ruin perturbative renormalizability because the more
powers of them that appear in a result, the more divergent that
result typically is.

For instance, a contribution that arises at $n$th order in
Newton's constant usually depends on $G$ through the dimensionless
combination $(G \Lambda^2)^n \propto (\ell_p/\delta)^{2n}$, where
$\Lambda \propto 1/\delta$ is the UV cutoff in momentum space
(equivalently, $\delta$ is the small-distance cutoff in position
space). By contrast, having more powers of dimensionless
couplings, or those having dimensions of inverse powers of length,
do not worsen UV divergences. Ever-worsening divergences ruin the
arguments that show for renormalizable theories that all
calculations are finite once a basic set of couplings are
appropriately redefined. Removal of divergences can be
accomplished, but only by introducing an infinite number of
coupling parameters to be renormalized.

Lack of renormalizability was for a long time regarded as a
fundamental obstacle to performing any quantum calculations within
gravity. After all, if every calculation is associated with a new
parameter that absorbs the new divergences, whose value must be
inferred experimentally, then there are as many parameters as
observables and no predictions are possible. If this were really
true, it would mean that any classical prediction of GR would come
with incalculable theoretical errors due to the uncontrolled size
of the quantum corrections. And the presence of such errors would
render meaningless any detailed comparisons between classical
predictions and observations, potentially ruining GR's
observational successes. How can meaningful calculations be made?

\subsection{Effective Field Theories}

As it happens, tools for making meaningful quantum calculations
using non-renormalizable theories exist, having been developed for
situations where quantum effects are more important than they
usually are for gravity [Weinberg 1979, Gasser 1984].

The key to understanding how to work with non-renormalizable
theories is to recognize that they can arise as approximations to
more fundamental, renormalizable physics, for which explicit
calculations are possible. The way non-renormalizable theories
arise in this case is as a low-energy/long-distance approximation
in situations for which short-distance physics is unimportant, and
so is coarse-grained or integrated out [Gell-Mann 1954, Wilson
1974].

For instance, consider the lagrangian density for the quantum
electrodynamics of electrons and muons
\begin{equation} \label{LQED}
 {\cal L}_{\scriptscriptstyle QED} = - \frac14 \, F_{\mu\nu}
 F^{\mu\nu} - \overline\psi (\gamma^\mu D_\mu + m) \psi
 - \overline\chi (\gamma^\mu D_\mu + M) \chi \,,
\end{equation}
where $m = m_e$ and $\psi$ (or $M = m_\mu \gg m_e$ and $\chi$) are
the electron (or muon) mass and field. Here $F_{\mu\nu} =
\partial_\mu A_\nu - \partial_\nu A_\mu$ and $D_\mu = \partial_\mu
+ ie A_\mu$, as usual, and $\gamma^\mu$ represents the Dirac
matrices --- that satisfy $\{\gamma^\mu , \gamma^\nu \} = 2
\eta^{\mu\nu} = 2 \hbox{diag}(-,+,+,+)$. This is a renormalizable
theory because all parameters, $e$, $m$ and $M$, have non-positive
dimension when regarded as a power of length in fundamental units.

Suppose now we choose to examine observables only involving the
electromagnetic interactions of electrons at energies $\omega \ll
M$ (such as the energy levels of atoms, for instance). Muons
should be largely irrelevant for these kinds of observables, but
not completely so. Muons are not completely irrelevant because
they can contribute to electron-photon processes at higher orders
in perturbation theory as virtual states.

It happens that any such effects due to virtual muons can be
described at low energies by the following {\em effective field
theory} of electrons and photons only:
\begin{eqnarray} \label{LeffQED}
 {\cal L}_{\rm\,eff} &=& - \frac14 \, F_{\mu\nu}
 F^{\mu\nu} - \overline\psi (\gamma^\mu D_\mu + m) \psi
 + \frac{k_1 \, \alpha}{30 \pi M^2} \, F^{\mu\nu} \Box F_{\mu\nu}
 + \cdots \\
 &=& - \frac14 \, F_{\mu\nu}
 F^{\mu\nu} - \overline\psi (\gamma^\mu D_\mu + m) \psi
 + \frac{k_1 \, \alpha }{15 \pi M^2} \, (\overline \psi \gamma_\mu \psi)
 (\overline \psi \gamma^\mu \psi)
 + \cdots \,, \nonumber
\end{eqnarray}
where the second line is obtained from the first by performing the
field redefinition
\begin{equation}
 A_\mu \to A_\mu + \frac{k_1 \,\alpha}{15\pi M^2} \, \Bigl[\Box A_\mu
 - ie \, (\overline \psi \gamma_\mu \psi) \Bigr] + \cdots \,.
\end{equation}
In both equations the ellipses describe terms suppressed by more
than two powers of $1/M$.

The lagrangian densities of eqs.~(\ref{LQED}) and (\ref{LeffQED})
are precisely equivalent in that they give precisely the same
results for {\em all}\, low-energy electron/photon observables,
provided one works only to leading order in $1/M^2$. If the
accuracy of the agreement is to be at the one-loop level, then
equivalence requires the choice $k_1 = 1$, and the effective
interaction captures the leading effects of a muon loop in the
vacuum polarization. If agreement is to be at the two-loop level,
then $k_1 = 1 + {\cal O}(\alpha)$ captures effects coming from
higher loops as well, and so on.

This example (and many many others) shows that it must be possible
to make sensible predictions using non-renormalizable theories.
This must be so because the lagrangian of eq.~(\ref{LeffQED}) is
not renormalizable --- its coupling has dimensions (length)${}^2$
--- yet it agrees precisely with the (very sensible) predictions
of QED, eq.~(\ref{LQED}). But it is important that this agreement
only works up to order $1/M^2$.

If we work beyond order $1/M^2$ in this expansion, we can still
find a lagrangian, ${\cal L}_{\rm\,eff}$, that captures all of the
effects of QED to the desired order. The corresponding lagrangian
requires more terms than in eq.~(\ref{LeffQED}), however, also
including terms like
\begin{equation}
 {\cal L}_4 = \frac{k_2 \, \alpha^2}{90 M^4}
 \, (F_{\mu\nu} F^{\mu\nu})^2 \,,
\end{equation}
that arise at order $1/M^4$. Agreement with QED in this case
requires $k_2 = 1 + {\cal O}(\alpha)$. Sensible predictions can be
extracted from non-renormalizable theories, but only if one is
careful to work only to a fixed order in the $1/M$ expansion.

What is useful about this process is that an effective theory like
(\ref{LeffQED}) is much easier to use than is the full theory
(\ref{LQED}). And any observable whatsoever may be computed once
the coefficients ($k_1$ and $k_2$ in the above examples) of the
various non-renormalizable interactions are identified. This can
be done by comparing its implications with those of the full
theory for a few specific observables.

What about the UV divergences associated with these new effective
interactions? They must be renormalized, and the many couplings
required to perform this renormalization correspond to the many
couplings that arise within the effective theory at successive
orders in $1/M$. But predictiveness is not lost because working to
fixed order in $1/M$ means that only a fixed number of effective
couplings are required in any given application.

At present this is the {\em only} known way to make sense of
perturbatively non-renormalizable theories. In particular it means
that there is a hidden approximation involved in the use of a
non-renormalizable theory --- the low-energy, $1/M$, expansion ---
that may not have been obvious from the get-go.

\subsection{GR as an effective theory}

What would this picture mean if applied to GR? First, it would
mean that GR must be regarded as the leading term in the
low-energy/long-distance approximation to some more fundamental
theory. Working beyond leading order would mean extending the
Einstein-Hilbert action to include higher powers of curvatures and
their derivatives, with the terms with the fewest derivatives
being expected to dominate at low energies [for a review see
Burgess 2004].

Since we do not know what the underlying theory is, we cannot hope
to compute the couplings in this effective theory from first
principles as was done above for QED. Instead we treat these
couplings as phenomenological, ultimately to be determined from
experiment.

The most general interactions involving the fewest curvatures and
derivatives, that are consistent with general covariance are
\begin{eqnarray}
\label{gravaction}
 - \, {{\cal L}_{\rm eff} \over \sqrt{- g}} &=& \lambda
 + \frac{M_p^2}{2}  \, R + a_1 \, R_{\mu\nu} \, R^{\mu\nu}
 \nonumber\\
 && \quad + a_2 \, R^2
 +  a_3 \, R_{\mu\nu\lambda\rho} R^{\mu\nu\lambda\rho}
 + a_4 \, \Box R \\
 &&\quad \quad + {b_1 \over m^2}\; R^3 + {b_2 \over m^2} \; R R_{\mu\nu}
 R^{\mu\nu} + {b_3\over m^2} \; R_{\mu\nu} R^{\nu\lambda}
 {R_\lambda}^\mu + \cdots \,, \nonumber
\end{eqnarray}
where ${R^\mu}_{\nu\lambda\rho}$ is the metric's Riemann tensor,
$R_{\mu\nu} = {R^\lambda}_{\mu\lambda\nu}$ is its Ricci tensor,
and $R = g^{\mu\nu}R_{\mu\nu}$ is the Ricci scalar, each of which
involves precisely two derivatives of the metric.

The first term in eq.~(\ref{gravaction}) is the cosmological
constant, which we drop because observations imply $\lambda$ is
(for some unknown reason, see below) extremely small. Once this is
done the leading term in the derivative expansion is the
Einstein-Hilbert action whose coefficient, $M_p = \left( {8\pi G}
\right)^{-1/2} = (\sqrt{8\pi} \; \ell_p)^{-1} \sim 10^{18}$ GeV,
has dimensions of mass (when $\hbar = c = 1$), and is set by the
value of Newton's constant. This is followed by curvature-squared
terms having dimensionless effective couplings, $a_i$, and
curvature-cubed terms with couplings inversely proportional to a
mass, $b_i/m^2$, (not all of which are written in
eq.~(\ref{gravaction})).

Although the numerical value of $M_p$ is known, the mass scale $m$
appearing in the curvature-cubed (and higher) terms is not. But
since it appears in the denominator it is the lowest mass scale to
have been integrated out that should be expected to dominate. What
its value should be depends on the scale of the applications one
has in mind. For applications to the solar system or to
astrophysics $m$ might reasonably be taken to be the electron
mass, $m_e$. But for applications to inflation, where the scales
of interest are much larger than $m_e$, $m$ would instead be taken
to be the lightest particle that is heavier than the scales of
inflationary interest.

\subsection{Power counting}

The Einstein-Hilbert term should dominate at low energies (since
it involves the fewest derivatives), and this expectation can be
made more precise by systematically identifying which interactions
contribute to a particular order in the semiclassical expansion.
To do so we expand the metric about an asymptotically static
background spacetime: $g_{\mu\nu} = \overline{g}_{\mu\nu} +
2h_{\mu\nu}/M_p$, and compute (say) the scattering amplitudes for
asymptotic graviton states that impinge onto the geometry from
afar.

If the energy, $\omega$, of the incoming states are all comparable
and similar to the curvatures scales of the background spacetime,
dimensional analysis can be used to give an estimate for the
energy-dependence of an $L$-loop contribution to a scattering
amplitude, ${\cal A}(\omega)$. Consider a contribution to this
amplitude that involves $E$ external lines and $V_{id}$ vertices
involving $d$ derivatives and $i$ attached graviton lines.
Dimensional analysis leads to the estimate:
\begin{equation}
\label{GRcount1a}
 {\cal A}(\omega) \sim \omega^2 M_p^2 \left( {1
 \over M_p} \right)^{E}
 \left( {\omega \over 4 \pi M_p} \right)^{2
 L} {\prod_{i} \prod_{d>2}} \left[{\omega^2 \over M_p^2}
 \left( {\omega \over m} \right)^{(d-4)}  \right]^{V_{id}} \,.
\end{equation}
Notice that no negative powers of $\omega$ appear here because
general covariance requires derivatives come in pairs, so the
index $d$ in the product runs over $d = 4 + 2k$, with $k =
0,1,2,...$.

This last expression displays the low-energy approximation alluded
to above because it shows that the small quantities controlling
the perturbative expansion are $\omega/M_p$ and $\omega/m$. Use of
this expansion (and in particular its leading, classical limit --
see below) presupposes both of these quantities to be small.
Notice also that because $m \ll M_p$, factors of $\omega/m$ are
much larger than factors of $\omega/M_p$, but because they do not
arise until curvature-cubed interactions are important, the
perturbative expansion always starts off with powers of
$\omega/M_p$.

\subsection{What justifies the classical approximation?}

Eq.~(\ref{GRcount1a}) answers a question that is not asked often
enough: What is the theoretical error made when treating
gravitational physics in the classical approximation? What makes
it so useful in this regard is that it quantifies the size of the
contribution to ${\cal A}(\omega)$ (or other observables) arising
both from quantum effects ({\it i.e.} loops, with $L \ge 1$), and
from terms normally not included in the lagrangian (such as
higher-curvature terms). This allows an estimate of the size of
the error that is made when such terms are not considered (as is
often the case).

In particular, eq.~(\ref{GRcount1a}) justifies why classical
calculations using GR work so well, and quantifies just how
accurate their quantum corrections are expected to be. To see
this, we ask which graphs dominate in the small-$\omega$ limit.
For any fixed process ({\it i.e.} fixed $E$) eq.~(\ref{GRcount1a})
shows the dominant contributions are those for which
$$ L = 0 \quad \hbox{and} \quad V_{id} = 0
 \;\; \hbox{for any} \;\; d > 2 \,. $$
That is, the dominant contribution comes from arbitrary tree
graphs constructed purely from the Einstein-Hilbert $(d=2)$
action. This is precisely the prediction of classical General
Relativity.

For instance, for the scattering of two gravitons about flat
space, $g(p_1) + g(p_2) \to g(p_1') + g(p_2')$, we have ${E} = 4$,
and eq.~(\ref{GRcount1a}) predicts the dominant energy-dependence
to be ${\cal A}(\omega) \propto (\omega/M_p)^2$. This is borne out
by explicit tree-level calculations [DeWitt 1967] which give
\begin{equation}
 {\cal A}_{\rm tree} = 8 \pi i G \,\left(
 \frac{s^3}{tu} \right)\,,
\end{equation}
for an appropriate choice of graviton polarizations. Here $s = -
(p_1 + p_2)^2$, $t = (p_1 - p_1')^2$ and $u = (p_1 - p_2')^2$ are
the usual Lorentz-invariant Mandelstam variables built from the
initial and final particle four-momenta, all of which are
proportional to $\omega^2$. This shows both that ${\cal A} \sim
(\omega/M_p)^2$ to leading order, and that it is the physical,
invariant, centre-of-mass energy, $\omega_{cm}$, that is the
relevant scale against which $m$ and $M_p$ should be compared.

The next-to-leading contributions, according to
eq.~(\ref{GRcount1a}), arise in one of two ways: either
\begin{eqnarray}
 && L = 1 \quad \hbox{and} \quad V_{id} = 0
 \;\; \hbox{for any} \;\; d > 2; \nonumber\\
 \hbox{or} &&
 L = 0, \; \sum_i V_{i4} = 1, \quad \hbox{and} \;\;
 V_{id} =0  \;\; \hbox{for} \;\; d > 4 \,. \nonumber
\end{eqnarray}
These correspond to one-loop (quantum) corrections computed only
using Einstein gravity; plus a tree-level contribution including
precisely one vertex from one of the curvature-squared
interactions (in addition to any number of interactions from the
Einstein-Hilbert term). The UV divergences arising in the first
type of contribution are absorbed into the coefficients of the
interactions appearing in the second type. Both are suppressed
compared to the leading, classical, term by a factor of
$(\omega/4\pi M_p)^2$. This estimate (plus logarithmic
complications due to infrared divergences) is also borne out by
explicit one-loop calculations about flat space [Weinberg 1965,
Dunbar 1995, Donoghue 1999].

This is the reasoning that shows why it makes sense to compute
quantum effects, like Hawking radiation or inflationary
fluctuations, within a gravitational context. For observables
located a distance $r$ away from a gravitating mass $M$, the
leading quantum corrections are predicted to be of order
$G\hbar/r^2c^3 = (\ell_p/r)^2$. For comparison, the size of
classical relativistic corrections is set by $2GM/rc^2 = r_s/r$,
where $r_s = 2GM/c^2$ denotes the Schwarzschild radius. At the
surface of the Sun this makes relativistic corrections of order
$GM_\odot/R_\odot c^2 \sim 10^{-6}$, while quantum corrections are
$G \hbar/R_\odot^2 c^3 \sim 10^{-88}$. Clearly the classical
approximation to GR is {\it extremely} good within the solar
system.

On the other hand, although relativistic effects cannot be
neglected near a black hole, since $2GM/r_s c^2 = 1$, the relative
size of quantum corrections near the event horizon is
\begin{equation}
 \left( \frac{\ell_p}{r_s} \right)^2 =
 \frac{G\hbar}{r_s^2c^3} =
 \frac{\hbar c}{4GM^2} \,,
\end{equation}
which is negligible provided $M \gg M_p$. Since $M_p$ is of order
tens of micrograms, this shows why quantum effects represent small
perturbations for any astrophysical black holes,\footnote{Small,
but not negligible, since the decrease in mass predicted by
Hawking radiation has no classical counterpart with which to
compete.} but would not be under control for any attempt to
interpret the gravitational field of an elementary particle (like
an electron) as giving rise to a black hole.

\subsection{Lessons learned}

What do these considerations tell us about how gravity behaves
over very small distances?

The good news is that it says that the observational successes of
GR are remarkably robust against the details of whatever
small-distance physics ultimately describes gravity over very
small distances. This is because {\em any} microscopic physics
that predicts the same symmetries (like Lorentz invariance) and
particle content (a massless spin-2 particle, or equivalently a
long-range force coupled to stress-energy) as GR, must be
described by a generally covariant effective action like
eq.~(\ref{gravaction}). Because this is dominated at low energies
by the Einstein-Hilbert action, it suffices to get the low-energy
particle content and symmetries right to get GR right in all of
its glorious detail [Deser 1970].

The bad news applies to those who think they know what the
fundamental theory of quantum gravity really is at small scales,
since whatever it is will be very hard to test experimentally.
This is because all theories that get the bare minimum right (like
a massless graviton), are likely to correctly capture all of the
successes of GR in one fell swoop. At low energies the only
difference between the predictions of {\em any} such theory is the
value of the coefficients, $a_i$ and $b_i$ {\em etc}, appearing in
the low-energy lagrangian (\ref{gravaction}), none of which are
yet observable.

There are two kinds of proposals that allow tests at low energies:
those that change the low-energy degrees of freedom (such as by
adding new light particles in addition to the graviton --- more
about these proposals below); and those that change the symmetries
predicted for the low-energy theory. Prominent amongst this latter
category are theories that postulate that gravity at short
distances breaks Lorentz or rotational invariance, perhaps because
spacetime becomes discrete at these scales.

At first sight, breaking Lorentz invariance at short distances
seems batty, due to the high accuracy with which experimental
tests verify the Lorentz-invariance of the vacuum within which we
live. How could the world we see appear so Lorentz invariant if it
is really not so deeper down? Surprisingly, experience with other
areas of physics suggests this may not be so crazy an idea; we
know of other, emergent, symmetries that can appear to be very
accurate at long distances even though they are badly broken at
short distances. Most notable among these is the symmetry
responsible for conservation of baryon number, which has long been
known to be an `accidental' symmetry of the Standard Model. This
means that for {\em any} microscopic theory whose low-energy
particle content is that of the SM, any violations of baryon
number must necessarily be described by a non-renormalizable
effective interaction [Weinberg 1979a, Wilczek 1979], and so be
suppressed by a power of a large inverse mass, $1/M$. This
suppression can be enough to agree with observations (like the
absence of proton decay) if $M$ is as large as $10^{16}$ GeV.

Could Lorentz invariance be similarly emergent? If so, it should
be possible to find effective field theories for which Lorentz
violation first arises suppressed by some power of a heavy scale,
$1/M$, even if Lorentz invariance is not imposed from the outset
as a symmetry of the theory. Unfortunately this seems hard to
achieve, since in the absence of Lorentz invariance it is
difficult\footnote{The situation would be different in Euclidean
signature, since then invariance under a lattice group of
rotations can suffice to imply invariance under $O(4)$
transformations, at least for the kinetic terms.} in an effective
theory to explain why the effective terms
\begin{equation}
 \partial_t \psi^* \partial_t \psi
 \qquad \hbox{and} \qquad
 \nabla \psi^* \cdot \nabla \psi \,,
\end{equation}
should have precisely the same coefficient in the low-energy
theory. (See however [Groot Nebbelink 2005] for some attempts.)
The problem is that the coefficients of these terms are
dimensionless in fundamental units, and so are unsuppressed by
powers of $1/M$. But the relative normalization of these two terms
governs the maximal speed of propagation of the corresponding
particle, and there are extremely good bounds (for some particles
better than a part in $10^{20}$) on how much this can differ from
the speed of light [see, for instance, Mattingly 2005 for a recent
review].

This underlines why proponents of any particular Quantum Gravity
proposal must work hard to provide the effective field theory
(EFT) that describes their low-energy limit [see Kostelecky 2004,
Mattingly 2005 for some gravitational examples]. Since all of the
observational implications are contained within the effective
theory, it is impossible to know without it whether or not the
proposal satisfies all of the existing experimental tests. This is
particularly true for proposals that claim to predict a few
specific low-energy effects that are potentially observable (such
as small violations of Lorentz invariance in cosmology). Even if
the predicted effects should be observed, the theory must also be
shown not to be in conflict with other relevant observations (such
as the absence of Lorentz invariance elsewhere), and this usually
requires an EFT formulation.

\section{Modifying Gravity over Long Distances}

There also has been considerable activity over recent years
investigating the possibility that GR might fail, but over very
long distances rather than short ones. This possibility is driven
most persuasively from cosmology, where the Hot Big Bang paradigm
has survived a host of detailed observational tests, but only if
the universe is pervaded by no less than {\em two} kinds of new
exotic forms of matter: dark matter (at present making up $\sim
25\%$ of the universal energy density) and Dark Energy (comprising
$\sim 70\%$ of the cosmic energy density). Because all of the
evidence for the existence of these comes from their gravitational
interactions, inferred using GR, the suspicion is that it might be
more economical to interpret instead the cosmological tests as
evidence that GR is failing over long distances.

But since the required modifications occur over long distances,
their discussion is performed most efficiently within an effective
lagrangian framework. These next paragraphs summarize my personal
take on what has been learnt to this point.

\subsection{Consistency issues}

An important consideration when trying to modify gravity over long
distances is the great difficulty in doing so in a consistent way.
Almost all modifications so far proposed run into trouble with
stability or unitarity, in that they predict unstable degrees of
freedom like `ghosts,' particles having negative kinetic energy.
The presence of ghosts in a low energy theory is generally
regarded as poison because it implies there are instabilities. At
the quantum level these instabilities usually undermine our
understanding of particle physics and the very stability of the
vacuum [see Cline 2004 for a calculation showing what can go
wrong], but even at the classical level they typically ruin the
agreement between the observed orbital decay of binary pulsars and
GR predictions for their energy loss into gravitational waves.

The origin of these difficulties seems to be the strong
consistency requirements that quantum mechanics and Lorentz
invariance impose on theories of massless particles having
spin-one or higher [Weinberg 1964, Deser 1970, Weinberg 1980],
with static (non-derivative) interactions. A variety of studies
indicate that a consistent description of particles with spins
$\ge 1$ always requires a local invariance, which in the cases of
spins 1, 3/2 and 2 corresponds to gauge invariance, supersymmetry
or general covariance, and this local symmetry strongly limits the
kinds of interactions that are possible.\footnote{The AdS/CFT
correspondence [Maldacena 1998] -- a remarkable equivalence
between asymptotically anti-de Sitter gravitational theories and
non-gravitational systems in one lower dimensions -- may provide a
loophole to some of these arguments, although its ultimate impact
is not yet known.} Although it remains an area of active research
[Dvali 2000], at present the only systems known to satisfy these
consistency constraints consist of relativistic theories of spins
0 through 1 coupled either to gravity or supergravity (possibly in
more than 4 spacetime dimensions).

\subsection{Dark Matter}

As might be expected, widespread acceptance of the existence of a
hitherto-unknown form of matter requires the concordance of
several independent lines of evidence, and this constrains one's
options when formulating a theory for dark matter. It is useful to
review this evidence when deciding whether it indicates a failure
of GR or a new form of matter.

The evidence for dark matter comes from measuring the amount of
matter in a region as indicated by how things gravitate towards
it, and comparing the result with the amount of matter that is
directly visible. Several types of independent comparisons
consistently point to there being more than 10 times as much dark,
gravitating material in space than is visible:\footnote{This is
consistent with the cosmological evidence that dark matter is
roughly 5 times more abundant than ordinary matter (baryons)
because most of the ordinary matter is also dark, and so is also
not visible.}
\begin{itemize}
\item {\it Galaxies:} The total mass in a galaxy may be inferred
from the orbital motion of stars and gas measured as a function of
distance from the galactic center. The results, for large galaxies
like the Milky Way, point to several times more matter than is
directly visible.
\item {\it Galaxy Clusters:} Similar measurements using the motion
of galaxies and temperature of hot gas in large galaxy clusters
also indicate the presence of much more mass than is visible.
\item {\it Structure Formation:} Present-day galaxies and galaxy
clusters formed through the gravitational amplification of
initially-small primordial density fluctuations. In this case the
evidence for dark matter arises from the interplay of two facts:
First, the initial density fluctuations are known to be very
small, $\delta \rho/\rho \sim 10^{-5}$, at the time when the CMB
was emitted. Second, small initial fluctuations cannot be
amplified by gravity until the epoch where non-relativistic matter
begins to dominate the total energy density. But this does not
give enough time for the initially-small fluctuations to form
galaxies unless there is much more matter present than can be
accounted for by baryons. The amount required agrees with the
amount inferred from the previous measures described above.
\end{itemize}

These in themselves do not show that the required dark matter need
be exotic, the evidence for which also comes from several sources
\begin{itemize}
\item {\it Primordial Nucleosynthesis:} The total mass density of
ordinary matter (baryons) in the universe can be inferred from the
predicted relative abundance of primordial nuclei created within
the Hot Big Bang. This predicted abundance agrees well with
observations, and relies on the competition between nuclear
reaction rates and the rate with which the universe cools. But
both of these rates themselves depend on the net abundance of
baryons in the universe: the nuclear reaction rates depend on the
number of baryons present; and the cooling rate depends on how
fast the universe expands, and so -- at least, in GR -- on its
total energy density. The success of the predictions of Big Bang
Nucleosynthesis (BBN) therefore fixes the fraction of the
universal energy density which can consist of baryons, and implies
that there can at most be a few times more baryons than what would
be inferred by counting those that are directly visible.
\item {\it The Cosmic Microwave Background (CMB):} CMB photons
provide an independent measure of the total baryon abundance. They
do so because sound waves in the baryon density that are present
when these photons were radiated are observable as small
temperature fluctuations. Since the sound-wave properties depend
on the density of baryons, a detailed understanding of the CMB
temperature spectrum allows the total baryon density to be
reconstructed. The result agrees with the BBN measure described
above.
\end{itemize}

There are two main options for explaining these observations.
Since dark matter is inferred gravitationally, perhaps the laws of
gravity differ on extra-galactic scales than in the solar system.
Alternatively, there could exists a cosmic abundance of a new type
of hitherto-undiscovered particle.

At present there are several reasons that make it more likely that
dark matter is explained by the presence of a new type of particle
than by changing GR on long distances. First, as mentioned above,
sensible modifications are difficult to make at long distances
that lack ghosts and other inconsistencies. Second, no
phenomenological modification of gravity has yet been proposed
that accounts for all the independent lines of evidence given
above (although there is a proposal that can explain the rotation
of galaxies [Milgrom 1983, Sanders 2002]).

On the other hand, all that is required to obtain dark matter as a
new form of matter is the existence of a new type of stable
elementary particle having a mass and couplings similar to those
of the $Z$ boson, which is already known to exist. $Z$ bosons
would be excellent dark matter candidates if only they did not
decay. A particle with mass and couplings like the $Z$ boson, but
which is stable --- called a Weakly Interacting Massive Particle
(WIMP) --- would naturally have a relic thermal abundance in the
Hot Big Bang that lies in the range observed for dark matter [for
a review, see Eidelman 2004]. New particles with these properties
are actually predicted by many current proposals for the new
physics that is likely to replace the Standard Model at energies
to be explored by the Large Hadron Collider (LHC).

At the present juncture the preponderance of evidence --- the
simplicity of the particle option and the difficulty of making a
modification to GR that works --- favours the interpretation of
cosmological evidence as pointing to the existence of a new type
of matter rather than a modification to the laws of gravity.

\subsection{Dark Energy}

The evidence for dark energy is more recent, and incomplete, than
that for dark matter. At present the evidence for its existence
comes from two independent lines of argument:

\begin{itemize}
\item {\it Universal Acceleration:} Since gravity is attractive,
one expects an expanding universe containing only ordinary (and
dark) matter and radiation to have a decelerating expansion rate.
Evidence for dark energy comes from measurements indicating the
universal expansion is {\em accelerating} rather than
decelerating, obtained by measuring the brightness of distant
supernovae [Perlmutter 1997, Riess 1997, Bahcall 1999]. According
to GR, accelerated expansion implies the universe is dominated by
something with an equation of state satisfying $p < - \rho/3$,
which is not true for ordinary matter, radiation or dark matter.
\item {\it Flatness of the universe:} An independent measure of
the dark energy comes from the observed temperature fluctuations
in the CMB. Because the CMB photons traverse the entire observable
universe before reaching us, their properties on arrival depend on
the geometry of the universe as a whole (and so also, according to
GR, on its total energy density). Agreement with observations
implies the total energy density is larger than the ordinary and
dark matter abundances, which fall short by an amount consistent
with the amount of dark energy required by the acceleration of the
universe's expansion [Komatsu 2009].
\end{itemize}

Again the theoretical options are the existence of a new form of
energy density, or a modification of GR at long distances.
Although there are phenomenological proposals for modifications
that can cause the universe to accelerate (such as [Dvali 2000]),
all of the previously described problems with long-distance
modifications to GR also apply here.

By contrast, there is a very simple energy density that does the
job, consisting simply of a cosmological constant --- {\em i.e.} a
constant $\lambda \simeq (3 \times 10^{-3}$ eV$){}^4$ in
eq.~(\ref{gravaction}), for which $p = - \rho$. This is
phenomenologically just what the doctor ordered, and agrees very
well with the observations.

The theoretical difficulty here is that a cosmological constant is
indistinguishable from the energy density of a Lorentz-invariant
vacuum,\footnote{The only known loophole to this arises if extra
dimensions exist, and are as large as 10 microns in size, because
in this case the vacuum energy can be localized in the extra
dimensions, and so curve these rather than the dimensions we see
[Arkani-Hamed 2000, Kachru 2000, Carroll 2003, Aghababaie 2004].
Whether this, together with supersymmetry, can solve the problem
is under active study [Burgess 2005].} since both contribute to
the stress tensor an amount $T_{\mu\nu} = \lambda \, g_{\mu\nu}$.
In principle, this should be a good thing because we believe we
can compute the vacuum energy. The problem is that ordinary
particles (like the electron) contribute such an enormous amount
--- the electron gives $\delta \lambda \simeq m_e^4 \simeq
(10^{6}$ eV$)^4$ --- that agreement with the observed value
requires a cancellation [Weinberg 1989] to better than one part in
$10^{36}$.

\subsection{Lessons learned}

Dark matter and dark energy are two forms of exotic matter, whose
existence is inferred purely from their gravitational influence on
visible objects. It is tempting to replace the need for two new
things with a single modification to gravity over very large
distances.

Yet the preponderance of evidence again argues against this point
of view. First, it is difficult to modify GR at long distances
without introducing pathologies. Second, it is difficult to find
modifications that account for more than one of the several
independent lines of evidence (particularly for dark matter). By
contrast, it is not difficult to make models of dark matter
(WIMPs) or dark energy (a cosmological constant). For dark energy
this point of view runs up against the cosmological constant
problem, which might indicate the presence of observably large
extra dimensions, but for which no consensus yet exists.

\section{Conclusions}

In summary, modifications to General Relativity are widely mooted
over both large and small distances. This chapter argues that
modifications at small distances are indeed very likely, and well
worth seeking. But unless the modification takes place just beyond
our present experimental reach ($\sim 10^{-19}$ m) [Arkani-Hamed
1998, Antoniadis 1998, Burgess 2005], it is also likely to be very
difficult to test experimentally. The basic obstruction is the
decoupling from long distances of short-distance physics, a
property most efficiently expressed using effective field theory
methods. The good news is that this means that the many
observational successes of GR are insensitive to the details of
whatever the modification proves to be.

Modifications to GR over very long distances are also possible,
and have been argued as more economical than requiring the
existence of two types of unknown forms of matter (dark matter and
dark energy). If so, consistency constraints seem to restrict the
possibilities to supplementing GR by other very light spin-0 or
spin-1 bosons (possibly in higher dimensions). The experimental
implications of such modifications are themselves best explored
using effective field theories. Unfortunately, no such a
modification has yet been found that accounts for all of the
evidence for dark matter or energy in a way that is both
consistent with other tests of GR and is more economical than the
proposals for dark matter or energy themselves.

To the extent that the utility of effective field theory relies on
decoupling, one might ask: What evidence do we have that
Planck-scale physics decouples? There are two lines of argument
that bear on this question. First, once specific modifications to
gravity are proposed it becomes possible to test whether
decoupling takes place. Perhaps the best example of a consistent
modification to gravity at short distances is string theory, and
all the present evidence points to decoupling holding in this
case. But more generally, if sub-Planckian scales do {\em not}
decouple, one must ask: Why has science made progress at all?
After all, although Nature comes to us with many scales,
decoupling is what ensures we don't need to understand them all at
once. If sub-Planckian physics does not decouple, what keeps it
from appearing everywhere, and destroying our hard-won
understanding of Nature?

\section*{Acknowledgements}

I thank the editors for their kind invitation to contribute to
this volume, and for their patience in awaiting my contribution.
My understanding of this topic was learned from Steven Weinberg,
who pioneered effective field theory techniques, and was among the
first to connect the dots explicitly about gravity's
interpretation as an effective field theory. My research is funded
by the Natural Sciences and Engineering Research Council of
Canada, as well as by funds from McMaster University and Perimeter
Institute.

\end{document}